\documentclass[aps,pre,twocolumn,showpacs,showkeys]{revtex4}
\usepackage{graphicx}%
\usepackage{dcolumn}
\usepackage{amsmath}   

\usepackage{bm}
\usepackage{longtable}

\begin{document}

\title{Terahertz response of dipolar impurities in polar liquids: On
  anomalous dielectric absorption of protein solutions
}
\author{Dmitry V.\ Matyushov}
\affiliation{Center for Biological Physics, Arizona State University, 
PO Box 871604, Tempe, AZ 85287-1604                   }
\email{dmitrym@asu.edu}
\begin{abstract}
  A theory of radiation absorption by dielectric mixtures is
  presented.  The coarse-grained formulation is based on the
  wavevector-dependent correlation functions of molecular dipoles of
  the host polar liquid and a density-density structure factor of the
  positions of the solutes. A nonlinear dependence of the absorption
  coefficient on the solute concentration is predicted and originates
  from the mutual polarization of the liquid surrounding the solutes
  by the collective field of the solute dipoles aligned along the
  radiation field. The theory is applied to terahertz absorption of
  hydrated saccharides and proteins. While the theory gives an
  excellent account of the observations for saccharides without
  additional assumptions and fitting parameters, experimental
  absorption coefficient of protein solutions significantly exceeds
  theoretical calculations within standard dielectric models and shows
  a peak against the protein concentration. A substantial polarization
  of protein's hydration shell is required to explain the differences
  between standard theories and experiment. When the correlation
  function of the total dipole moment of the protein with its
  hydration shell from numerical simulations is used in the present
  analytical model an absorption peak similar to that seen is
  experiment is obtained. The result is sensitive to the specifics of
  protein-protein interactions in solution. Numerical testing of the
  theory requires the combination of terahertz dielectric and
  small-angle scattering measurements.
\end{abstract}

\maketitle

\section{Introduction}
\label{sec:1}
Dielectric spectroscopy of mixtures is a well-established technique
which requires theoretical modeling for the data interpretation.  The
models of dielectric response of mixtures traditionally operate by
assuming that a mixture can be separated into macroscopic dielectric
bodies.  Among the commonly used models are the Maxwell-Wagner theory
\cite{Scaife:98} and various formulations of the effective-medium
approximation \cite{Choi:99}. Both assume that a dielectric constant
can be assigned to each component, and the latter also requires that
the physical properties of the host and the impurity are not
dramatically different.

The recent rapid development of dielectric techniques to study
mixtures \cite{Takashima:89}, in particular in the terahertz (THz)
frequency window \cite{Beard:02}, aims at a different
length-scale. The interest is mainly driven by the desire to learn
about electrostatics of nano-scale objects, such as biopolymers
\cite{Yokoyama:01,Bergner:05,Zhang:06,Xu:06,Knab:07,Ebbinghaus:08,Frauenfelder:09},
nano-crystals \cite{Baxter:06}, and nano-confined fluids
\cite{Alcoutlabi:05}. In particular, one hopes that the properties of
the nano-scale interface between the solvent and the solute can be
effectively probed by the dielectric response. This goal is
complicated by the fact that essentially any relaxation event linked
to electrical dipoles in the system contributes to the integral
experimental signal, and theory is required to separate different
components. While fully atomistic models will be the ultimate goal of
the theory, it is still useful to develop coarse-grained approaches
employing the length-scale intermediate between macroscopic dimensions
of classical theories \cite{Scaife:98,Choi:99} and fully atomistic
length-scale.

This paper presents a coarse-grained model of the dielectric response
of dipolar mixtures, aiming in particular at the THz frequency
window. The model does not assume that solutes can be described as
dielectric bodies, neither does it assume dielectric continuum for a
polar solvent. The polar liquid is characterized by its
wavevector-dependent correlation functions \cite{DMjcp1:04}, and a
similar approach is invoked for the solutes characterized by their
density structure factor. However, instead of using completely
atomistic structures, the solutes are modeled by effective spheres
characterized by dipole moments, polarizabilities, effective radii,
etc. The assumption of solute sphericity does not pose a fundamental
restriction on the theory since it can be extended to solutes of
non-spherical shapes made by overlapping vdW spheres of the composing
atoms \cite{DMjcp2:08}. However, this simplification allows us to come
up with a set of compact analytical equations applicable to analyzing
experimental data.

The theory is applied to the analysis of the absorption coefficient of
THz radiation. Recent measurements on hydrated saccharides
\cite{Heyden:08} and proteins \cite{Ebbinghaus:07} have shown
qualitatively different types of dependencies of THz dielectric
absorption on concentrations of these two types of solutes. The
current theory gives an excellent account of the observations on
saccharides, but fails to reproduce the protein experiments when the
dipole moment of the protein is assigned to the solute. It is
suggested that hydrated proteins introduce solvation electrostatics
qualitatively different from the dielectric response of typical
dipolar mixtures \cite{DMpre2:08}. Specifically, hydration layers
nearest to the protein (ca.\ 15 \AA{} in thickness) become polarized and
thus carry a significant dipole moment with the relaxation dynamics
different from that of the protein \cite{DMjpcb2:09}.  This ``elastic
ferroelectric bag'' \cite{DMpre2:08} surrounding the protein
significantly enhances the effective dipole moment of the solute
observed on the large wavelength of THz radiation and can account for
the observed anomalous dielectric absorption of protein solutions
\cite{Ebbinghaus:07}. Since both the scenario of the rigid protein
dipole and the dipole dressed by the ferroelectric bag can be
introduced into the formalism, the present theory provides a tool to
separate this new physics from what can be described within the
traditional understanding of dipolar liquids and solvation
electrostatics.

\section{Dielectric response of mixtures}
\label{sec:2}
We consider a polar liquid with dipolar impurities (solutes). The
impurities are larger than the molecules of the host liquid in most
cases of practical interest and can physically be realized as
molecules or small colloids (nanoparticles).  The physics of the
problem is clearly presented by separating the process of inserting
the impurities into two steps: (i) the creation of a hard-core cavity
in the liquid and (ii) the polarization of the host polar liquid by
the partial charges of the overall neutral solute.  We note in passing
that the restriction of the neutral solute can be lifted when ionic
conductivity is not an issue, such as the case for many THz dielectric
measurements.

\begin{figure}
  \centering
  \includegraphics*[width=5cm]{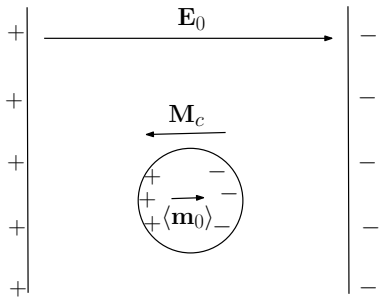}
  \caption{Schematic diagram of a conventional dielectric impedance
    experiment. The electric field $\mathbf{E}_0$ (in the absence of
    dielectric) is perpendicular to the plane of the liquid film such
    that the field in the dielectric is $\mathbf{E}_0/ \epsilon$. This is a
    longitudinal field as it sets up the direction of symmetry
    breaking in the homogeneous liquid. The polarization of the cavity
    in the liquid induces the depolarization field and the cavity
    dipole $\mathbf{M}_c$ opposite to the direction of the external
    field. The average solute dipole $\langle \mathbf{m}_0\rangle$, aligned along
    the external field, enhances both the dielectric response and the
    depolarization field of the empty cavity.  }
  \label{fig:1}
\end{figure}

The creation of a cavity in a polar liquid results, in terms of
standard dielectric theories \cite{Landau8,Boettcher:73}, in a
depolarization field, i.e.\ charges on the cavity's surface that
create the cavity dipole moment $\mathbf{M}_c$ opposite to the
direction of the external field.  In the standard setup of the
dielectric spectroscopy experiment shown in Fig.\ \ref{fig:1} the
electric field is longitudinal, i.e.\ parallel to the direction of
breaking the isotropic symmetry of the liquid by an external
perturbation. The dipole of a spherical cavity of volume $\Omega_0$ is then
\cite{Boettcher:73} $\mathbf{M}_c = - 3\mathbf{P}^L\Omega_0 /(2\epsilon +1)$,
where $\epsilon$ is the dielectric constant of the homogeneous liquid and
$\mathbf{P}^L$ is the longitudinal (superscript ``L'') polarization
field created by the external source of the electric field
$\mathbf{E}_0$. Since $\mathbf{P}^L=(\epsilon -1)\mathbf{E}_0/(4\pi\epsilon)$, the
cavity dipole decreases with increasing $\epsilon$. Standard low-frequency
(high $\epsilon$) dielectric measurements of polar liquids are therefore
fairly insensitive to impurities.

The dipole moment of the solute orients itself in the external field
amplifying the dielectric response. This effect is partially
compensated by an additional polarization of the cavity surface by the
internal dipole acting to enhance the cavity dipole in the direction
opposite to the external field (Fig.\ \ref{fig:1}). The solute dipoles
can be considered as independent in the limit of infinite dilution,
and the change of the dielectric response is linear in the dipoles'
concentration. This approximation limits the range of concentrations
by the requirement that the Onsager radius of the solute-solute
dipolar interactions is below the average distance between them.

The situation becomes more complex for a finite concentration of
solute dipoles when an additional effect of their collective field
gains in importance. The alignment of solute dipoles in the external
field creates a net average dipole moment $\langle\mathbf{m}_0\rangle$ (Fig.\
\ref{fig:1}) and a corresponding non-zero net electric field that can
potentially polarize cavities and alter their cavity dipoles. Since
internal fields are commonly large compared to the external field,
this effect, nonlinear in the solutes' concentration, can be
potentially significant.

The arguments we have presented so far apply to the standard
dielectric impedance measurements employing longitudinal electric
fields. THz experiments employ a different geometry where the
absorption of a pulse of electromagnetic wave propagating orthogonally
to a thin (ca.\ 100 $\mu$m) film is measured
\cite{Bergner:05,Knab:07}. In this case, the electric field is
transversal, i.e.\ it is perpendicular to the direction of axial
symmetry breaking introduced in the isotropic liquid by the direction
of the wavevector \cite{Madden:84,Neumann:86}. One measures then the
transverse dielectric response and the cavity is polarized
differently. The dipole moment of the cavity along the field becomes
$\mathbf{M}_c^{T}= - \mathbf{P}^T\Omega_0 \times 3\epsilon/(2\epsilon +1)$, where the
transverse polarization (superscript ``T'') is $\mathbf{P}^T=(\epsilon
-1)\mathbf{E}_0/(4\pi)$. It is clear that the cavity dipole produced in
response to the transversal field is not screened by the high
dielectric constant of a polar liquid. Microwave absorption
measurements are therefore expected to be significantly more sensitive
to impurities than conventional dielectric measurements. This
distinction is the physical basis of the sensitivity of the
transversal absorption experiments to electrostatic changes in
molecular or nano-scale solutes \cite{Beard:02}.

\section{Response function}
We now turn our attention to a detailed analysis of the transverse
dielectric response of dipolar mixtures. In order to approach this
problem we will use the approximation of linear response of the
solvent to the electric field of the solute. The linear response
approximation states that the solvent response function is insensitive
to the magnitude of the solute electric field and in fact can be
calculated for a fictitious solute with all partial charges turned off
(zero dipole for a dipolar solute) \cite{DMjcp2:08}. Even though the
electrostatic response is linear, the response to the solute repulsive
core cannot be calculated within linear models since the repulsive
potential of the solute produces a large and nonlinear perturbation of
the solvent structure. This perturbation renormalizes the spectrum of
the solvent fluctuations modifying the linear (Gaussian) response
function \cite{Chandler:93}. In dielectric theories, this modification
is included by imposing boundary conditions on the solution of the
Poisson equation. The problem becomes way more complex at the
molecular level and is commonly solved in terms of angular-dependent
distribution functions \cite{Hansen:03}.

We will adopt here Chandler's formulation of the Gaussian model
\cite{Chandler:93} in which the linear response function, modified by
the presence of solute, is sought by imposing the condition of
vanishing solvent density from the solute's hard core. In case of
polarization response, this condition implies the polarization field
$\mathbf{P}$ vanishing from the hard core of the solute. One can then
define a generating functional of the polarization field as follows
\cite{DMjcp1:04}
\begin{equation}
  \label{eq:5}
  \begin{split}
  \mathcal{G}[\mathbf{E}_0] = &\int \exp\left[-(\beta/2)
    \mathbf{P}*\bm{\chi}_s^{-1}*\mathbf{P} + \beta \mathbf{E}_0*\mathbf{P}
  \right] \\ & \prod_{i,\Omega_0}
  \delta\left[\mathbf{P}(\mathbf{r})\right]\mathcal{D}\mathbf{P} .   
  \end{split}
\end{equation}
Here, $\mathbf{E}_0$ is an external electric field, the asterisk
denotes both the volume integration and tensor contraction, and
$\beta=1/(k_{\text{B}}T)$ is the inverse temperature. Further, $\bm{\chi}_s$
is the 2-rank tensor of the Gaussian fluctuations of the polarization
field in the homogeneous solvent and the product of delta functions
runs over all points within solute's hard-core of volume $\Omega_0$ and
over all solutes (index $i$). This term ensures that the polarization
field vanishes from the volume of each solute in the mixture.

Functional derivatives of $\mathcal{G}[\mathbf{E}_0]$ over the
external field $\mathbf{E}_0$ produce correlation functions of the
polarization field of the solvent in the presence of solutes.  The
Gaussian integral over the polarization field $\mathbf{P}(\mathbf{r})$
can be calculated exactly resulting in a Gaussian functional in the
external field $\mathbf{E}_0$. The corresponding renormalized response
function $\bm{\chi}$ gains most compact representation in the inverted
$\mathbf{k}$-space \cite{DMjcp1:04}. It can be written in the
$\mathbf{k},\omega$-representation in the following form
\begin{equation}
  \label{eq:6}
  \begin{split}
  \bm{\chi}&(\mathbf{k}_1,\mathbf{k}_2,\omega)= 
  \bm{\chi}_s(\mathbf{k}_1,\omega)\delta_{\mathbf{k}_1,\mathbf{k}_2} \\
          & - \sum_i \bm{\chi}^{R}(\mathbf{k}_1,\omega) \cdot  
e^{i(\mathbf{k}_1-\mathbf{k}_2)\cdot\mathbf{r}_i} 
\theta_0(\mathbf{k}_1-\mathbf{k}_2) \cdot \bm{\chi}_s(\mathbf{k}_2,\omega) . 
\end{split}
\end{equation}
Here, $\delta_{\mathbf{k}_1,\mathbf{k}_2}= (2\pi)^3
\delta(\mathbf{k}_1-\mathbf{k}_2)$ and $\theta_0(\mathbf{k})$ is the Fourier
transform of the step function defining the excluded volume of the
solute the translational dynamics of which are neglected. The
direct-space Heaviside function $\theta_0(\mathbf{r})$ is equal to unity
within the solute and is equal to zero outside the solute. The
inverted-space function is given by the Fourier transform
\begin{equation}
  \label{eq:7}
  \theta_0(\mathbf{k}) = \int_{\Omega_0} e^{i\mathbf{k}\cdot\mathbf{r}}
  d\mathbf{r} ,
\end{equation}
where integration is over the solute volume $\Omega_0$. 

The response function of the mixture
$\bm{\chi}(\mathbf{k}_1,\mathbf{k}_2,\omega)$ depends on two wavevectors
$\mathbf{k}_1$ and $\mathbf{k}_2$ separately, instead of
$\mathbf{k}_1-\mathbf{k}_2$ of the homogeneous liquid, because of the
inhomogeneous response produced by each solute marked by index
$i$. This response function combines the dipolar response function of
the homogeneous liquid $\bm{\chi}_s(\mathbf{k},\omega)$, the information about
the solute shape incorporated into $\theta_0(\mathbf{k})$, and the
renormalized function $\bm{\chi}^{R}(\mathbf{k},\omega)$ (see below).

The response function of an axially-symmetric dipolar liquid is
expandable into longitudinal (L) and transverse (T) projections
\cite{Wertheim:71,Hansen:03}
\begin{equation}
  \label{eq:8}
  \bm{\chi}_s(\mathbf{k},\omega) = \chi^L(k,\omega)\mathbf{J}^L +
    \chi^T(k,\omega)\mathbf{J}^T,
\end{equation}
where $\mathbf{J}^L=\mathbf{\hat k}\mathbf{\hat k}$ and $\mathbf{J}^T
= \mathbf{1} - \mathbf{\hat k}\mathbf{\hat k}$ are the orthogonal longitudinal
and transverse dyads.  The $k=0$ values of the response projections
are directly related to the frequency-dependent dielectric constant of
the host liquid
\begin{equation}
  \label{eq:9}
  \begin{split}
    4\pi\chi_s^L(0,\omega) & = 1 - \epsilon (\omega)^{-1}, \\
    4\pi \chi_s^T(0,\omega) & = \epsilon (\omega) -1 .
  \end{split}
\end{equation}
The entire $k,\omega$-dependence of the projections $\chi^{L,T}(k,\omega)$ is given
in Ref.\ \onlinecite{DMjcp1:05}, but only the transverse projection is
required for the problem considered here (see below).

The last function in Eq.\ (\ref{eq:6}) that requires definition is
$\bm{\chi}^R(\mathbf{k},\omega)$. This function appears in the solution for
the generating functional in Eq.\ (\ref{eq:5}) as a result of
renormalizing the dipolar response of the homogeneous liquid by the
solute cavity. It thus contains the information about both the solvent
and the solute \cite{DMjcp1:04,DMjcp1:05}. Only $k=0$ transverse
projection of this function appears in the equations for the
transverse dielectric response of the dipolar mixture and that is
given by the following equation
\begin{equation}
  \label{eq:11}
  \chi^{R,T}(0,\omega) = \frac{3\epsilon (\omega)}{2\epsilon (\omega)+1} . 
\end{equation}

We will now use Eq.\ (\ref{eq:6}) to calculate the transverse dipole
moment $M^T(\omega)$ of the dielectric sample produced in response to the
electric field of the electromagnetic radiation oscillating with
frequency $\omega$
\begin{equation}
  \label{eq:12}
\mathbf{E}_0(t) = \mathbf{\hat e}^T E_0e^{i\omega t} .
\end{equation}
Here, the polarization unit vector $\mathbf{\hat e}^T$ is
perpendicular to the direction of propagation $\mathbf{k}$. 

The dipole moment $M^T(\omega)$ combines two contributions: the dipole
moment induced directly by the external field of the radiation
(radiation wave-length is much larger than any molecular scales in the
system) and an additional collective polarization induced by the
solute dipoles aligned along the external field. These two
contributions are described by correspondingly the first and the
second summands in the following relation
\begin{equation}
  \label{eq:13}
  \begin{split}
  M^T(\omega)&=\mathbf{\hat e}^T\cdot \bm{\chi}(0,0,\omega)\cdot\mathbf{\hat e}^T E_0 \\
        &+ \mathbf{\hat e}^T\cdot \bm{\chi}(0,\mathbf{k},\omega) * \sum_i
        \mathbf{T}(\mathbf{k})
             e^{i\mathbf{k}\cdot\mathbf{r}_i} \cdot
               \mathbf{m}_{0,i}(\omega) ,
 \end{split} 
\end{equation}
where $\mathbf{T}(\mathbf{k})$ is the Fourier transform of the dipolar
tensor $\mathbf{T} = - \nabla_{\mathbf{r}}\nabla_{\mathbf{r}'} |\mathbf{r} -
\mathbf{r}'|^{-1}$ and, as above, the asterisk refers to integration
over $\mathbf{k}$-space and tensor contraction. In addition,
$\mathbf{m}_{0i}(\omega)$ is the solute's dipole moment aligning itself
with the oscillating external field.

The dipole moment $\mathbf{m}_{0i}(\omega)$ is a sum of two components: the
electronic dipole induced instantaneously (on the time-scales of
interest) by the external filed and a permanent dipole inertially
rotated by the torque imposed by the external field. The inertial
component can be calculated from the linear-response approximation
\cite{Hansen:03} with the result
\begin{equation}
  \label{eq:14}
  \mathbf{\hat e}^T\cdot \mathbf{m}_{0i}(\omega) = \left(\alpha_{0e} + \alpha_{0,n}^T[1 - i\omega\Phi(-\omega)] \right)f_d(\omega)E_0. 
\end{equation}
Here, $\alpha_{0,e}$ is the solute electronic polarizability and the
permanent dipole polarizability is given as
\begin{equation}
  \label{eq:10}
  \alpha_{0,n}^T = (\beta m_0^2/2)g_{0,\text{K}}^T,
\end{equation}
where $g_{0,\text{K}}^T$ is the transverse Kirkwood factor
\cite{SPH:81} of the correlated orientations of the solute dipoles:
\begin{equation}
  \label{eq:15}
  g_{0,\text{K}}^T = \left\langle \sum_j \left[\mathbf{\hat e}_i \cdot
      \mathbf{\hat e}_j - (\mathbf{\hat e}_i \cdot \mathbf{\hat k}) 
      (\mathbf{\hat k} \cdot \mathbf{\hat e}_j)\right] \right\rangle .
\end{equation}
Here, $\mathbf{\hat e}_j$ are the unit vectors of the solute dipoles.
If the dielectric constant $\epsilon_0$ can be assigned to the solutes, then
$\epsilon_0 = 1 + 2\pi \beta m_0^2 \rho_0 g_{0,\text{K}}^T$, $\rho_0=N_0/V$.  We also
assumed isotropic polarizability of the solute and, in addition, for
solution problems, the permanent dipole $m_0$ should be properly
renormalized from the gas-phase value by the effect of the solute
polarizability \cite{Onsager:36,Boettcher:73,DMepl:08}. Further, the
factor $f_d(\omega)$ in Eq.\ (\ref{eq:14}) is the Onsager directing field
correction \cite{Onsager:36,Boettcher:73} accounting for the
difference between the electric field of the radiation and the local
electric field imposing torque on the solute dipole.

The Laplace-Fourier transform $\Phi(\omega)$ in Eq.\ (\ref{eq:14}) represents
correlated rotational dynamics of the solute dipole
\begin{equation}
  \label{eq:25}
  \Phi(\omega) = (m_0^2 g_{0,\text{K}}^T)^{-1} \int_0^{\infty}
  \mathbf{\hat e}^T\cdot \langle\mathbf{m}_{0}(t) \mathbf{M}_0(0)\rangle \cdot \mathbf{\hat
    e}^T e^{i\omega t} dt, 
\end{equation}
where $\mathbf{M}_0=\sum_j \mathbf{m}_{0,j}$ is the total solute
dipole in the sample. In case of a single-time Debye rotational
relaxation with the relaxation time $\tau_0$ the term in the square
brackets in Eq.\ (\ref{eq:14}) gains the form
\begin{equation}
  \label{eq:26}
  1 - i\omega\Phi(-\omega)= (1 + i\omega\tau_0)^{-1} .
\end{equation}

The Debye approximation in Eq.\ (\ref{eq:26}) is typically sufficient
for rigid dipoles dissolved in a polar solvent. The situation
potentially becomes more complex for soft nano-scale solutes,
biopolymers in the first place. The dynamics of the dipole moment is
then affected by low-frequency vibrations \cite{Knab:07} altering
$\Phi(-\omega)$. As we discuss below, the inclusion of a non-vanishing dipole
moment of the protein's hydration shell, with its own dynamics, makes
the problem even more non-trivial, further complicating the form of
$\Phi(-\omega)$.

The first term in Eq.\ (\ref{eq:13}) can be easily calculated by
combining Eqs.\ (\ref{eq:9}) and (\ref{eq:11}) and noting that $\theta_0(0)
= \Omega_0$ [Eq.\ (\ref{eq:7})]. This calculation then results in a simple
relation for the difference between the response function of the
mixture $\chi_{\text{mix}}(\omega)=M^T(\omega)/(VE_0)$ and of the homogeneous
liquid $4\pi \chi(\omega)=\epsilon(\omega)-1$
\begin{equation}
  \label{eq:16}
  4\pi \Delta \chi  =  - \eta_0 f(\omega) .
\end{equation}
Here, $\Delta \chi = \chi_{\text{mix}}(\omega) -\chi(\omega) $, $\eta_0 = N_0 \Omega_0/ V$ is the
volume fraction of the solutes in the mixture with the overall volume
$V$, and
\begin{equation}
  \label{eq:17}
  f(\omega) = \frac{3\epsilon (\omega) ( \epsilon (\omega) -1 )}{2 \epsilon (\omega) +1} .
\end{equation}

We note here that a more simple (and elegant) derivation of the
response function of a low-concentration mixture as given by Eqs.\
(\ref{eq:16}) and (\ref{eq:17}) can be found in Ref.\
\onlinecite{Landau8}.  Equations (\ref{eq:16}) and (\ref{eq:17}) also
represent a low-concentration limit of the Maxwell-Wagner formula
\cite{Scaife:98}. Our microscopic consideration is thus consistent
with macroscopic arguments. The microscopic description is however
required to get correctly the second summand in Eq.\ (\ref{eq:13})
describing the collective response of an ensemble of solute
dipoles. This is what we consider next.

The response function of the solvent to the presence of the solute
includes two parts corresponding to two summands in Eq.\
(\ref{eq:6}). The first summand represents the response of the liquid
to an infinitely small solute which does not perturb the spectrum of
dipolar fluctuations of the liquid. This part is easy to calculate and
its relative contribution to the response is $\Delta \chi (\omega)/ \chi(\omega) = y_0(\omega)
f_d(\omega)$, where
\begin{equation}
  \label{eq:23}
  y_0(\omega)=(4\pi/3)\rho_0\alpha_{0,e} + (2\pi/3) \beta g_{0,\text{K}}^T m_0^2 \rho_0 \left[1 
  - i\omega \Phi(-\omega) \right]
\end{equation}
is the dipolar density of solutes defined in analogy with a similar
quantity of homogeneous liquids \cite{Scaife:98}.

The contribution from the second term in Eq.\ (\ref{eq:6}) is the
correction of the solvent response introduced by the excluded volume
of the solute. This calculation is more complex. After some algebra
one arrives at the mixture susceptibility relative to the
susceptibility of the homogeneous liquid
\begin{equation}
  \label{eq:19}
  \begin{split}
   \Delta \chi(\omega)/ \chi(\omega) &= - \eta_0  \frac{3\epsilon(\omega)}{2\epsilon(\omega)+1} \\ 
                &+ y_0(\omega) f_d(\omega)
   \left(1 - \frac{3\epsilon(\omega)}{2\epsilon(\omega)+1} I(\omega,\eta_0,R) \right) . 
  \end{split}
\end{equation}

The only non-trivial part in this equation is the integral $
I(\omega,\eta_0,R)$, arising from the combined effect of the volume excluded by
the solute from the solvent, many-body solute-solute correlations, and
microscopic correlations between the dipoles of the solvent. It is
given by the relation
\begin{equation}
  \label{eq:20}
   I(\omega,\eta_0,R) = \frac{6R}{\pi} \frac{\epsilon(0)-1}{\epsilon(\omega)-1} \int_0^{\infty} 
   dk j_1^2(kR) S_0(k,\eta_0)\frac{\chi_s^T(k,\omega)}{\chi_s^T(0,0)} ,
\end{equation}
in which $j_1(x)$ is the first-order spherical Bessel function and
$R=(\sigma_0 + \sigma)/2$ is the distance of the closest approach of the water
molecules with the effective hard-sphere diameter $\sigma$ to the solute
characterized by its hard-sphere diameter $\sigma_0$.

The density-density structure factor $S_0(k,\eta_0)$ in Eq.\
(\ref{eq:20}) is responsible for a nonlinear dependence of the
response function of the mixture on the solute concentration. The
$k=0$ value of the structure factor $S_0(0,\eta_0)$ ($S_0(0,\eta_0)\to 1$ at
$\eta_0\to 0$) is the reduced compressibility of the solute component of
the mixture. It is equal to the experimentally measurable osmotic
compressibility \cite{Kirkwood:51,Belloni:00}
\begin{equation}
  \label{eq:29}
  S_0(0,\eta_0) = \chi_{\text{osm}} = \left( \partial \rho_0 \over \partial (\beta \Pi)
  \right)_{\text{osm}} , 
\end{equation}
where $\Pi$ is the osmotic pressure and the derivative is taken under
the condition of osmotic equilibrium.

\begin{figure}
  \centering
  \includegraphics*[width=7cm]{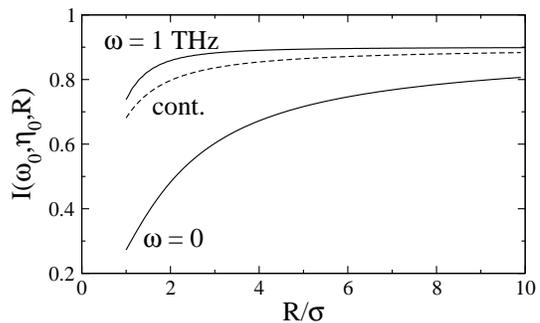}
  \caption{$I(\omega,\eta_0,R)$ calculated at $\omega=0$ and $\omega =
    1$ THz as indicated in the plot vs the reduced distance of closest
    solvent approach to the solute $R/ \sigma = (\sigma_0 / \sigma +1 )/2$.  The
    dashed line is the dielectric-continuum result of Eq.\
    (\ref{eq:27}); $\eta_0 = 0.1$ and the solvent parameters are those
    of water (see Appendix).  }
  \label{fig:2}
\end{figure}

The transverse dipolar correlation function $\chi_s^T(k,\omega)$ in Eq.\
(\ref{eq:20}) does not depend on the solute concentration, but
incorporates spacial transverse correlations between dipoles in the
polar liquid.  We provide its functional form here for completeness
and refer the reader to Refs.\ \onlinecite{DMjcp1:05} and
\onlinecite{DMjcp2:04} for a more detailed account of this problem
\begin{equation}
  \label{eq:24}
  \frac{\chi_s^T(0,0)}{\chi_s^T(k,\omega)} = \frac{S^T(k)}{S^T(0)} + 
   \frac{1}{1+p'(k\sigma)^2} \left(\frac{\epsilon(0)-1}{\epsilon(\omega)-1} -1\right). 
\end{equation}
In this equation, $S^T(k)$ is the static structure factor of
transverse dipolar fluctuations. A simple extension of the
mean-spherical solution for dipolar fluids \cite{Wertheim:71} gives
$S^T(k)$ consistent with numerical simulations \cite{DMjcp2:04}.  This
formalism is used here for numerical calculations of the function
$I(\omega,\eta_0)$ in Eq.\ (\ref{eq:20}). Finally, the parameter $p'$ in Eq.\
(\ref{eq:24}) quantifies the relative contribution of translational vs
rotational motions of liquid's dipoles in the overall response as
discussed in Ref.\ \onlinecite{Bagchi:91}.

The approximation of continuous dielectric corresponds to the neglect
of the $\mathbf{k}$-dependence in the transverse response function
$\chi_s^T(k,\omega)$ in Eq.\ (\ref{eq:20}) assuming $\chi_s^T(k,\omega)\simeq
\chi_s^T(0,\omega)$. The dependence of frequency then disappears from the
integral $I(\omega,\eta_0)$ which simplifies to
\begin{equation}
  \label{eq:27}
  I(\eta_0,R) = (6R/\pi) \int_0^{\infty}  dk j_1^2(kR) S_0(k,\eta_0) .
\end{equation}

The dielectric-continuum integral $I(\eta_0,R)$ is equal to unity for an
ideal solution when $S(k,0)=1$. This ideal-solution/continuum limit
then results in a simple equation for the mixture's dielectric
response
\begin{equation}
  \label{eq:28}
   \Delta \chi(\omega)/ \chi(\omega) 
    = - \eta_0  \frac{3\epsilon(\omega)}{2\epsilon(\omega)+1} - y_0(\omega) f_d(\omega) \frac{\epsilon(\omega)-1}{2\epsilon(\omega)+1} .
\end{equation}
It shows that the presence of very dilute solute dipoles lowers the
transverse response because an enhanced depolarization of the cavity
wins over the direct alignment of the solute dipoles by the external
field. It is clear that this result cannot sustain itself as the
concentration of dipolar impurities grows since the limit of a
negative dielectric constant can potentially be reached. Solution
non-ideality must slow the negative decay of the mixture
susceptibility or change its sign to positive.

The continuum integral $I(\eta_0,R)$ can be rewritten in
$\mathbf{r}$-space as
\begin{equation}
  \label{eq:51}
  I(\eta_0,R) = 1 + (\rho_0/\Omega_0)\int d\mathbf{r}_1d\mathbf{r}_2
  f_0(\mathbf{r}_1) h_0(\mathbf{r}_{12}) f_0(\mathbf{r}_2) , 
\end{equation}
where $h_0(\mathbf{r}_{12})$,
$\mathbf{r}_{12}=\mathbf{r}_2-\mathbf{r}_1$ is the pair correlation
function of the solutes and $f_0(\mathbf{r})$ are Mayer $f$-functions
representing hard cores of the solutes. If long-ranged interactions
between the solutes are neglected, the lowest-order density expansion
of the pair correlation function $h_0(\mathbf{r}_{12})$ yields the
third virial coefficient $C_{112}$ of the mixture of hard spheres of
diameter $R$ (components 1) and diameter $\sigma_0$ (component 2):
$I=1-(3\rho_0/ \Omega_0) C_{112}$, $\Omega_0 = (4\pi/3)R^3$. The third virial
coefficient of the hard-sphere mixture is known
\cite{Kihara:75,Barrio:99}. For solutes much larger than solvent, one
can put $R \simeq \sigma_0/2$ with the result $I(\eta_0) = 1 - (\eta_0/8)(3 + 65/24)$.
This simple equation compares reasonably well with the direct
numerical integration using the Percus-Yevick (PY) density structure
factor. The numerical integrals can be approximated by a polynomial of
$R/ \sigma_0$ and $\eta_0$, and this fit is provided in Appendix for $0\leq \eta_0 \leq
0.3$.

The interactions between hydrated proteins are complex, and the
structure factor from hard-sphere repulsions can be used only for a
limited range of ionic strengths when Coulomb forces are sufficiently
screened \cite{Zhang:07,Shukla:08}.  The structure factor $S_0(k,\eta_0)$
is directly measured by small-angle scattering \cite{Liu:05,Shukla:08}
and can be numerically reconstructed from a linear combination of a
repulsive and attractive potentials; a combination of Yukawa
potentials is often used \cite{Liu:05}. The small-$k$ part of the
structure factor is strongly affected by long-range interactions, and
there is a peak at $q_m\simeq 2\pi\sqrt[3]{n_0}$ at the average distance
between the solutes in solution. Since the amplitude of the peak is
relatively small \cite{Shukla:08}, a general insight into how
correlations between hydrated proteins affect the dielectric response
can be gained from an empirical approximation for $S_0(k,\eta_0)$. The
following approximation (analogous to the empty core model
\cite{Croxton:75}) follows directly from the low-density expansion of
the direct correlation function of a hard sphere
\begin{equation}
  \label{eq:30}
   S_0(k,\eta_0)= \left[1+a j_1(k\sigma_0)/(k\sigma_0) \right]^{-1}
\end{equation}
in which the constant $a$ is chosen to reproduce the osmotic
compressibility $a=3(S_0(0,\eta_0)^{-1} -1)$.  The resulting integral is
just a function of $a$ ($R \simeq \sigma_0/2$). Its numerical value can be
approximated by a Pad\'e form, $I(a)= (1 + 0.0908308 a - 0.00226567
a^2)/(1 + 0.131266 a - 0.00434023 a^2)$, which allows one to use the
osmotic compressibility, affected by both repulsions and attractions,
as input to obtain the dielectric response. The approximation in Eq.\
(\ref{eq:30}) is accurate up to $\eta_0\simeq0.1$ when compared to the direct
integration with the PY density structure factor.
  
It is worth noting at this point that the continuum approximation is
inaccurate at low frequencies $\omega \simeq 0$ overestimating the cavity
polarization in the entire range of solute sizes of common interest
(Fig.\ \ref{fig:2}).  This happens because of a very sharp decay of
the structure factor $S^T(k)$ at small $k$-values \cite{DMjcp2:04}
which, in the continuum limit, is replaced by its $k=0$ value
$S^T(0)$.  The continuum approximation becomes more accurate as the
frequency increases and the dielectric constant drops (Fig.\
\ref{fig:2}), but it needs to be tested before applied in a frequency
range of interest. Nevertheless, in the range of THz frequencies, the
continuum limit [Eq.\ (\ref{eq:27})] presents a useful simplification
of Eq.\ (\ref{eq:20}), which, in conjunction with Eq.\ (\ref{eq:30}),
yields the dielectric response solely in terms of observable
quantities.

\begin{figure}
  \centering
  \includegraphics*[width=7cm]{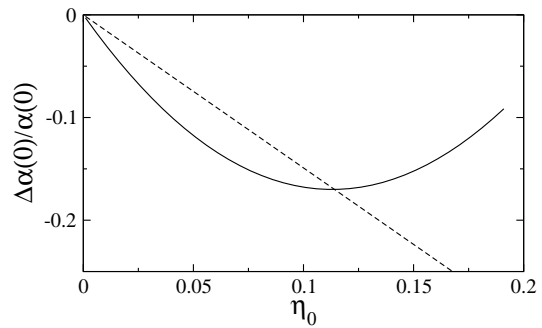}
  \caption{Relative change in the absorption coefficient of the
    mixture at $\omega=0$ [see Eqs.\ (\ref{eq:2}) and (\ref{eq:3})] as a
    function of the volume fraction of the solute $\eta_0$.  The solid
    line refers to the entire response function from Eq.\
    (\ref{eq:19}), while the dashed line shows the contribution of the
    first term only. The latter, linear in $\eta_0$, is the limit of zero
    solute dipole. The overall nonlinear dependence on $\eta_0$ is the
    result of mutual polarization of the cavities by the solute
    dipoles. The solute size and dipole are those of the $\lambda_{6-85}$
    protein. }
  \label{fig:3}
\end{figure}

The actual dependence of the dielectric response on the solute volume
fraction is more complex than a nearly linear decay suggested by Eq.\
(\ref{eq:28}). It is shown in Fig.\ \ref{fig:3} where a static $\omega=0$
response is calculated for parameters specific to $\lambda_{5-86}$ protein
discussed below. Baxter's solution of the PY closure \cite{Hansen:03}
for $S_0(k,\eta_0)$ was used in these calculations. In addition, the
microscopic transverse response function of the solvent dipoles was
taken according to Ref.\ \onlinecite{DMjcp1:05} and the static
structure factor was calculated from a corrected mean-spherical
approximation suggested in Ref.\ \onlinecite{DMjcp2:04}. The
dependence of the dielectric response on $\eta_0$ is curved down, thus
eliminating the dielectric catastrophe following from the linear
extrapolation of Eq.\ (\ref{eq:28}). However, the shape of the
concentration dependence depends on frequency, and the curvature is
just the opposite one for the THz response (see below).

A notion regarding theory's approximations is relevant here.  One
might argue that the point-dipole model is too restrictive for the
electrostatic field of a protein with typically a non-zero overall
charge and the prevalence of charged residues on its surface. We
believe that the approximations adopted here are adequate, and the
theory might actually be more quantitative than it seems. First, the
solvent response function is independent of the solute charge in the
linear response approximation \cite{DMjcp2:04} and is identical to the
one obtained for a fictitious solute with all charges turned off. The
linear response approximation might obviously fail, and that certainly
puts a restriction on the current theory. Second, the perturbation
Hamiltonian for the current problem is the interaction of the sample
dipole moment with the external electric field of the radiation. Since
the THz wavelength obviously exceeds any molecular dimension, a
dipolar approximation is appropriate for solutes of nano-scale
dimension. Finally, the total solute charge can contribute to
conductivity \cite{Rudas2:06} that is normally subtracted from the
dielectric response and is insignificant in the THz frequency range.
The dipole moment of charged solutes is then defined relative to the
solute's center of mass \cite{Rudas2:06}.

\section{Comparison to experiment}
\label{sec:3}
One of parameters reported in THz dielectric measurements is the
relative absorption coefficient $\Delta \alpha(\omega) / \alpha (\omega)$, where $\Delta
\alpha(\omega)=\alpha_{\text{mix}}(\omega) - \alpha(\omega) $ is the change in absorption
coefficient of the mixture relative to the pure liquid. The absorption
coefficient is defined \cite{WangCom:85,McQuarrie:00} as the ratio of
the rate of energy dissipation by the medium $\langle \mathcal{\dot E}\rangle_{\omega}$
over the Poynting vector $S(\omega)$ of the incident radiation
\begin{equation}
  \label{eq:1}
  \alpha(\omega) = \frac{\langle\mathcal{\dot E}\rangle_{\omega} }{S(\omega)} .
\end{equation}
By combining the standard equations for the Poynting vector in
dielectric media \cite{Landau8,McQuarrie:00} with energy dissipation
in terms of the dielectric response function $\chi(\omega)$ one gets the
equation
\begin{equation}
  \label{eq:2}
  \alpha(\omega) = \frac{4\pi\omega}{c} \frac{\chi''(\omega)}{\sqrt{\epsilon'(\omega)}} 
\end{equation}
which can be applied either to the mixture or to the pure liquid ($c$
is the speed of light in vacuum). 

Assuming that the deviation of the response $\Delta \chi(\omega)$ caused by
impurities is small compared to the dielectric response of the pure
liquid, one can easily derive an expression for the relative change of
the absorption coefficient
\begin{equation}
  \label{eq:3}
  \frac{\Delta \alpha(\omega)}{\alpha (\omega)} = \frac{4 \pi \Delta
    \chi''(\omega)}{\epsilon ''(\omega)} - \frac{2\pi \Delta
    \chi'(\omega)}{\epsilon' (\omega)} .
\end{equation}
In this equation, the variation of both the imaginary and the real
parts of the response are taken into account when impurities are
introduced into the polar liquid. In particular, for solutes with
small dipole moment, one can drop the term proportional to $y_0(\omega)$ in
Eq.\ (\ref{eq:19}) and arrive at a simple relation
\begin{equation}
  \label{eq:4}
   \frac{\Delta \alpha(\omega)}{\alpha (\omega)} = -\eta_0 \left[\frac{f''(\omega)}{\epsilon''(\omega)} - \frac{ f'(\omega)}{2\epsilon'(\omega)} 
   \right],
\end{equation}
where $f(\omega)$ is given by Eq.\ (\ref{eq:17}).

\begin{figure}
  \centering
  \includegraphics*[width=7cm]{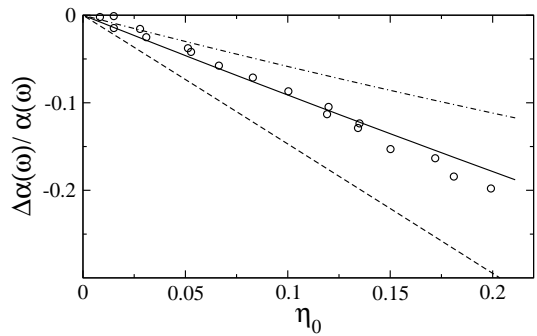}
  \caption{$\Delta\alpha(\omega)/ \alpha(\omega)$ at $\omega=2.5$ THz calculated from Eq.\
    (\ref{eq:4}) (solid line) and measured experimentally
    \cite{Heyden:08} (points) for the aqueous solution of
    trehalose. The dashed and dash-dotted lines represent,
    correspondingly, contributions from the first and second terms in
    Eq.\ (\ref{eq:4}) such that the solid line is their
    difference. The frequency-dependent dielectric constant of water
    in the THz range of frequencies was taken from Ref.\
    \onlinecite{Yada:09} and the molecular volume of trehalose $\Omega_0 =
    278$ \AA$^3$ \cite{Gharsallaoui:08} was used to convert from
    experimentally reported molar concentrations to volume fractions
    (see Appendix for the details of calculations).  }
  \label{fig:4}
\end{figure}

Figure \ref{fig:4} shows the comparison of Eq.\ (\ref{eq:4}) (lines)
with the experimental dependence (points) of the absorption
coefficient on the concentration of trehalose dissolved in liquid
water \cite{Heyden:08}. The details of the calculations and the
parameters used to produce the plot are given in the Appendix. Because
of the small dipole moment of trehalose, a complete calculation of the
dielectric response function of the mixture is not required (the term
proportional to $y_0(\omega)$ in Eq.\ (\ref{eq:19}) is small) and Eq.\
(\ref{eq:4}) is sufficient. The dashed and dash-dotted lines in Fig.\
\ref{fig:4} show the first (imaginary part) and second (real part)
terms in Eq.\ (\ref{eq:4}). It is clear that changes in the imaginary
and real parts of the dielectric susceptibility upon the addition of
impurities are comparable in magnitude and should both be included.
The only solute parameter entering Eq.\ (\ref{eq:4}) is its
volume. Equation Eq.\ (\ref{eq:4}) can therefore be used to determine
molecular volumes of weakly polar solutes by means of dielectric
measurements.

\begin{figure}
  \centering
  \includegraphics*[width=7cm]{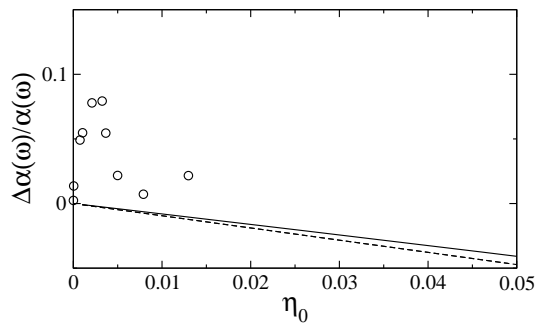}
  \caption{Relative change in the absorption coefficient at 2.25 THz
    mimicking the solution of protein $\lambda_{6-85}$ studied by THz
    spectroscopy in Ref.\ \onlinecite{Ebbinghaus:07}. The solid line
    is the dielectric response calculated from Eq.\ (\ref{eq:19}) with
    $m_0=61$ D ($g_{0,\text{K}}^T=2/3$) and the effective radius of
    12.1 \AA{} and the points are experimental measurements
    \cite{Ebbinghaus:07}.  The dashed line refers to the first term in
    Eq.\ (\ref{eq:19}) representing the polarization of the solute
    cavities by the external electric field. }
  \label{fig:5}
\end{figure}

In an attempt to see what might be the theory prediction for the case
of protein solutions we have mimicked the conditions reported in Ref.\
\onlinecite{Ebbinghaus:07} where the absorbance of the solution of a
five helix bundle protein $\lambda_{6-85}$ \cite{YangNature:03}
showed a maximum at the volume fraction of protein below 1\% (points in
Fig.\ \ref{fig:5}).  The calculations (see Appendix for the parameters
used) show almost no effect of proteins' dipoles and a negative
contribution to the absorption, as in the case of trehalose above and
in an obvious disagreement with the experiment.

There is also a clear difference between Figs.\ \ref{fig:3} and
\ref{fig:5}. While Fig.\ \ref{fig:3} shows a clear effect of the
mutual cavity polarization by solutes' dipolar fields for the same set
of parameters, there is almost no effect of the solute dipolar
component in Fig.\ \ref{fig:5} (cf.\ solid and dashed lines). The
difference comes from the dynamical effect. The solute dipoles do not
have time to reorient on the time-scale of the THz pulse and the
corresponding contribution is strongly diminished by the relaxation
$1/(\omega \tau_0)$ term. The THz pulse thus probes almost exclusively the
electronic polarizability of the solvated proteins.

For the solute dipoles to be seen in the THz response, either a much
faster relaxation or a significantly larger effective dipole are
required. Faster relaxation of protein's dipole seems improbable given
that numerical simulations show an almost exclusively single-component
rotational relaxation with the relaxation time in the range 3--6 ns
\cite{Rudas2:06}.  The hydration shell thus emerges as the most
probable candidate to explain the differences between the theory and
experiment.

\begin{figure}
  \centering
  \includegraphics*[width=8cm]{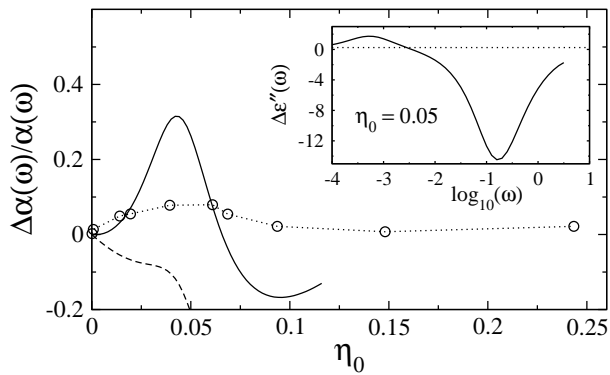}
  \caption{Change in absorption coefficient of solution of
    plastocyanin relative to bulk water at 2.25 THz. Dipole
    correlation function of the protein with the surrounding water
    shell was taken from MD simulations \cite{DMjpcb2:09}. The solute
    volume fraction $\eta_0$ was calculated by adding the width of the
    hydration shell ($20$ \AA) to the radius of plastocyanin (16.8 \AA).
    The points represent the experimental results for $\lambda_{6-85}$
    protein \cite{Ebbinghaus:07} recalculated from experimental molar
    concentrations by using the combined volume of plastocyanin and
    its polarized water shell. The solid curve refers to the
    calculations done with the hard-sphere solute-solute structure
    factor in Eq.\ (\ref{eq:20}). In order to show the sensitivity of
    the results to the solute-solute correlations, the dashed line
    represents the continuum integral $I(\eta_0)=1-3\eta_0$ with the slope
    against $\eta_0$ much exceeding that for hard-sphere solutes. The
    inset shows the change in dielectric loss of the plastocyanin
    solution at $\eta_0=0.05$ relative to bulk water against frequency
    measured in 10$^{12}$ s$^{-1}$. }
  \label{fig:6}
\end{figure}

In order to obtain more quantitative insights into the problem,
results of numerical simulations of protein solutions are required.
We found recently \cite{DMpre2:08,DMjpcb2:09} that, in accord with the
suggested interpretation of experimental THz data
\cite{Ebbinghaus:07}, proteins are capable of polarizing their
hydration shells $\simeq 10-15$ \AA{} into bulk water. This polarization
results in a significant non-zero average dipole moment of the
hydration shell $\langle |\mathbf{m}_w |\rangle$, which reached the value of $\simeq
10^3$ D in simulations of metalloprotein plastocyanin
\cite{DMpre2:08}. The dynamics of this ferroelectric cluster around
the protein are however decoupled from a much slower tumbling of the
protein occurring on the time-scale of nanoseconds. The relaxation of
the shell's dipole $\mathbf{m}_w$ is clearly two-component, with a
very fast initial relaxation on a sub-picosecond time-scale, followed
by a low-amplitude tail lasting hundreds of picoseconds. The fast
component correlates with low-frequency vibrations of the protein
deforming water's ``elastic ferroelectric bag'' \cite{DMjpcb2:09}.

In this picture, the solute dipole $\mathbf{m}_0$ should be replaced
with the sum of protein's and shell's dipoles $\mathbf{M}=\mathbf{m}_0
+\mathbf{m}_w$. The dynamics of this total dipole gives input to
determine function $\Phi(\omega)$, which, together with $\langle M^2\rangle$, yields
$y_0(\omega)$ [Eq.\ (\ref{eq:23})]. These parameters were extracted from
simulations of plastocyanin carrying the negative charge of $-8$ in
its oxidized state and hydrated by $N_w=21076$ TIP3P waters
\cite{DMpre2:08,DMjpcb2:09}.  The shell of water molecules of width 20
\AA{} was added to the effective radius of the protein to obtain the
effective radius of the protein/water cluster and the volume fraction
of coupled protein/water dipoles in solution (see Appendix for
details). The dielectric response of the solution was then calculated
from Eqs.\ (\ref{eq:19}) and (\ref{eq:2}).

Figure \ref{fig:6} shows the concentration dependence of the solution
absorption coefficient with the PY hard-sphere structure factor
$S_0(k,\eta_0)$ (solid line). The points, shown for reference, are
data on $\lambda_{6-85}$ protein \cite{Ebbinghaus:07} rescaled with the
volume of the plastocyanin/water cluster.  The calculation indeed
yields a maximum in the absorption coefficient which turns to negative
values with increasing volume fraction. The outcome of these
calculations is sensitive to the form of the density structure factor
and, therefore, to protein-protein interactions in solution. In order
to illustrate this point, the dashed line in Fig.\ \ref{fig:6} shows
the result of calculations with a stronger effect of repulsions and
thus a steeper decay of $S_0(0,\eta_0)$ with increasing $\eta_0$.

The hard-sphere model might not be adequate for all proteins and
electrolytes. For instance, for the ionic strength employed in Ref.\
\cite{Ebbinghaus:07} (0.05 M), the interactions between hydrated
bovine serum albumin (BSA) proteins are dominated by electrostatic
repulsions \cite{Zhang:08}. These proteins are negatively charged,
similarly to plastocyanin, and the long-range interactions are
dominated by the screened Coulomb potential. The osmotic
compressibility $S(0,\eta_0)$ of BSA quickly drops with increasing
protein concentration to the level $S(0,\eta_0)\simeq 0.1-0.2$ and then does
not significantly change when the concentration is further increased
\cite{Zhang:08}. With such a dependence of $S(0,\eta_0)$ on the volume
fraction $\eta_0$ the peak in absorption vanishes (Fig.\
\ref{fig:6}). Note that no absorption peak against protein
concentration was detected for BSA in dielectric terahertz
measurements at $\omega = 1.56$ THz \cite{Xu:06}.

The inset in Fig.\ \ref{fig:6} shows the frequency dependence of the
dielectric loss $\Delta \epsilon''(\omega)$. As is seen, the change of the loss
relative to bulk water can be either positive or negative, depending
on the frequency range. A complex concentration dependence seen for
the absorption coefficient in Fig.\ \ref{fig:6} is the cumulative
effect of the concentration dependencies of $\epsilon'_{\text{mix}}(\omega)$ and
$\epsilon''_{\text{mix}}(\omega)$.

\begin{figure}
  \centering
  \includegraphics*[width=8cm]{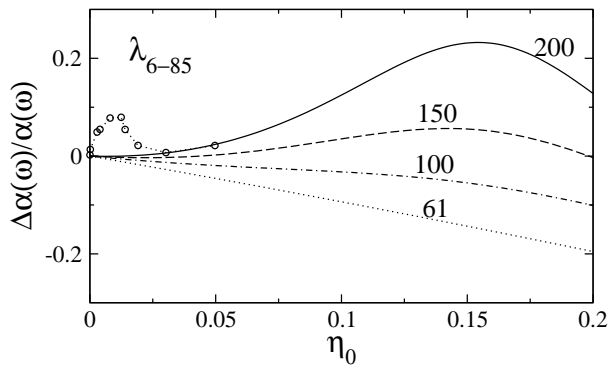}
  \caption{Change in absorption coefficient of solution of $\lambda_{6-85}$
    protein at 2.25 THz relative to bulk water. The normalized dipole
    correlation function of the protein with the surrounding water
    shell $\Phi(\omega)$ was taken from MD simulations of plastocyanin
    metalloprotein \cite{DMjpcb2:09}. The density structure factor of
    the proteins was constructed from the hard-sphere and Yukawa
    effective potential [Eq.\ (\ref{eq:18})] obtained in Ref.\
    \cite{Kim:08} by fitting small-angle scattering data. The curves
    refer to different dipole moments (in D) of the protein-water
    cluster as indicated in the plot. The cluster dipoles were assumed
    uncorrelated, $g_{\text{K}}^T=2/3$. The points are the
    experimental results from Ref.\ \cite{Ebbinghaus:07} converted to
    volume fraction with the hard-sphere diameter $\sigma_0=37.8$ \AA{} from
    the effective protein-protein interaction potential \cite{Kim:08}.
  }
  \label{fig:7}
\end{figure}

The $\lambda_{6-85}$ protein is uncharged and the corresponding
protein-protein interaction can be modelled either as a sum of soft
repulsion and exponentially decaying attraction or, alternatively, as
a sum of hard-sphere ($u_{\text{HS}}$) and attractive Yukawa
potentials \cite{Kim:08}:
\begin{equation}
  \label{eq:18}
  u(r) = u_{\text{HS}}(r) -\epsilon \left(\sigma_0/r\right) 
                         e^{-(r - \sigma_0)/ \delta} \theta(r-\sigma_0) . 
\end{equation}
In Fig.\ \ref{fig:7} we used this latter approximation for the
interaction potential to calculate $S_0(k,\eta_0)$ \cite{Javid:07} and
then applied this structure factor to the calculation of the THz
absorption coefficient. In the absence of dipole moment dynamics for
this protein, we used the normalized self-correlation function of the
protein-water dipole from plastocyanin simulations
\cite{DMpre2:08,DMjpcb2:09}. A set of curves in Fig.\ \ref{fig:7}
refer to different values of the dipole moment of the protein-water
cluster, with the lowest curve corresponding to the protein dipole
alone. Qualitatively, the absorption curves do go through maxima with
increasing dipole of the solute, and the protein solution absorbs
stronger than bulk water.  However, the maxima are broader than in
experiment and the agreement is only qualitative at best.

\section{Discussion}
The present model of the dielectric response targets physical
situations when large solutes dissolved in polar solvents do not
extend to dimensions of a dielectric material.  Large cavities in
polar liquids carry depolarization dipoles oriented oppositely to the
external field, with their magnitudes scaling linearly with the solute
volume. These depolarization dipoles accumulate a negative
contribution to the absorption coefficient. The intrinsic solute
dipoles, which align along the external electric field, increase the
absorption and also produce a non-zero local electric field that
re-polarizes neighboring cavities. This collective effect, non-linear
in the solute concentration, is sensitive to the solute-solute
correlations and is described by convoluting the solvent dipolar
response with the density structure factor of the dissolved solutes.

This model performs exceptionally well when tested against
experimental THz measurements for weakly polar impurities (Fig.\
\ref{fig:4}). In this case, only depolarization of cavities
contributes to the response, and that part of the problem seems to be
well captured by dielectric theories.  Even though solvation of
saccharides distorts the structure of water on the microscopic scale
\cite{Weingartner:01,Lee:05} and slows down the dynamics of the
hydration layer \cite{Paolantoni:09}, THz absorption seems to be
insensitive to such changes, and the resulting signal is well
described by a purely dielectric response. This conclusion is
consistent with the recent light scattering spectra of trehalose
solutions \cite{Paolantoni:09} suggesting only a local perturbation of
the water structure restricted to the first solvation shell, which is
typical for many small molecular solutes.

Polar impurities introduce both the effect of individual solute
dipoles and their collective polarization effect.  The
response-function formalism employed here does not involve any
large-scale changes in the solvent structure induced by the
solute. This formulation then fails to reproduce the anomalous
increase in the absorption of protein solutions over that of bulk
water \cite{Zhang:06,Ebbinghaus:07}. Computer simulations
\cite{DMpre2:08} show instead a high extent of cooperativity between
hydration shells and protein's motions. In addition, a significant
polarization of the water shell extending 10--20 \AA{} from the protein
surface into the bulk is observed. When the magnitude and correlation
function of the protein-water total dipole are substituted into the
equations for the solution response, the theory shows a maximum in the
absorption coefficient qualitatively similar to experimental
observations.  The maximum can therefore be considered as an
observable signature of the ``elastic ferroelectric bag'' found by
simulations \cite{DMpre2:08}.  The shape of this anomalous absorption
maximum is however sensitive to the interprotein interaction potential
and will be affected by several factors including protein's ionization
state and the ionic strength of the solution.

\acknowledgments This research was supported by the National Science
Foundation (CHE-0910905). The author is grateful to David Leitner for
sharing the simulation results on the lambda repressor protein and to
David LeBard for his help with the plastocyanin data.

\appendix

\section{Details of calculations}
\label{appB}
The dependence of the absorption coefficient on frequency arises
predominantly from the frequency-dependent dielectric constant of the
solvent. Dielectric measurements of water \cite{Yada:09} in the THz
range, extended to more typical low-frequency dielectric values, have
been used to produce Figs.\ \ref{fig:2} and \ref{fig:4}--
\ref{fig:7}. The dielectric constant is given by the following
relation \cite{Yada:09}
\begin{equation}
  \label{eqA:1}
  \epsilon(\omega) = \frac{\Delta \epsilon_1}{1-i\omega\tau_1} + \frac{\Delta
    \epsilon_2}{1-i\omega\tau_2} + \frac{A_S}{\omega_s^2-\omega^2 -
    i\omega \gamma_s } + \epsilon_{\infty},
\end{equation}
where $\Delta \epsilon_1 = 73.9$, $\Delta \epsilon_2 =1.56$, and $\epsilon_{\infty}=2.34$. The Debye
relaxation times and the parameters of the resonant component are:
$\tau_1=8.76$ ps, $\tau_2=0.224$ ps, $\omega_s/2\pi = 5.3$ THz, $\gamma_s/2\pi = 5.30$
THz, $A_S/(2\pi)^2=35.1$ THz$^2$. The parameter $f_d(\omega)$ [Eq.\
(\ref{eq:14})] accounts for the difference between the external and
the local directing (torque) fields. It depends on frequency through
the dielectric constant. This parameter is often associated with the
field within an empty cavity in a liquid \cite{Boettcher:73}. An
expression recently derived by us for this property \cite{DMepl:08}
was used in the calculations: $f_d(\omega) = [7(\epsilon(\omega)+1)^2 +
8\epsilon(\omega)]/[12\epsilon(\omega)(2\epsilon(\omega)+1)]$.

Since the polarizability of many organic substances is close to
$\alpha_{0,e}=\sigma_0^3/16$, the parameter of dipolar density of the
solutes [Eq.\ (\ref{eq:23})] was taken in the form
\begin{equation}
  \label{eqA:2}
  y_0(\omega) = \left[\frac{1}{2} +
     4g_{0,\text{K}}^T (m_0^*)^2\left(1 - i\omega \Phi(-\omega)
     \right) 
     \right] \eta_0 ,
\end{equation}
where $(m_0^*)^2= \beta \langle \mathbf{M}^2 \rangle / \sigma_0^3$ is a reduced effective
dipole, $\mathbf{M}$ is the entire dipole moment of the protein-water
cluster. 

Simulations of hydrated plastocyanin were reported previously
\cite{DMpre2:08}. The presently used data \cite{DMjpcb2:09} represent
the same simulation protocol applied to the oxidized (total charge
$-8$) state of plastocyanin extended to a larger number of waters in
the simulation box, $N_w = 21076$.  For plastocyanin calculations
$\mathbf{M}$ represents the total dipole of the protein and water
shell extending 20 \AA{} from the protein surface into the bulk. This
latter magnitude was added to the effective radius of the protein
listed in Table \ref{tab:1} to obtain the effective radius of the
water/protein cluster. The averaged square of the protein/water dipole
calculated from the simulation trajectory was $\langle \mathbf{M}^2\rangle =1.44 \times
10^6$ D$^2$. The response function $\Phi(\omega)$ was obtained as a
Laplace-Fourier transform of the three-exponent fit of the simulated
correlation function
\begin{equation}
  \label{eqA:5}
  \Phi(t) = \sum_{i=1}^3 A_i e^{-t/ \tau_i},
\end{equation}
where $A_i=\{0.84,0.11,0.05\}$ and $\tau_i=\{0.14,1790,6.3 \}$ ps. 

Other solute parameters used in the calculations are listed in Table
\ref{tab:1}, the hard sphere diameter of water was taken at the value
of $\sigma=2.87$ \AA, and the inertial parameter $p'$ in Eq.\ (\ref{eq:24})
was set at the value of $p'=0.1$ \cite{DMjcp1:05}.  The rotational
relaxation times of the solutes were taken at $\tau_0=50$ ps for
trehalose and $\tau_0\simeq 3$ ns for the two proteins. The former number is
consistent with the second relaxation process extracted from the
dielectric response and simulations of hydrated saccharides
\cite{Weingartner:01}, while the latter is typical for rotational
dynamics of proteins \cite{Rudas2:06}.

\begin{table}
\centering
\caption{\label{tab:1} Solute parameters used in the calculations.}
 \begin{ruledtabular} 
\begin{tabular}{lccc}
Solute &  $(\sigma_0/2)$/ \AA{} & $m_0$/D & $\tau_0$/ns \\
\hline\\ 
Trehalose & 8.2 & 1.75 & 0.05\\
$\lambda_{6-85}$ & 12.1\footnotemark[1] & 61\footnotemark[2] & 3 \\
Plastocyanin\footnotemark[3] & 16.8\footnotemark[4] & 248\footnotemark[5] & 
            2.8\footnotemark[6] \\
\end{tabular}
 \end{ruledtabular}
 \footnotetext[1]{From Ref.\ \cite{Ebbinghaus:07}. The following set
   of parameters from Ref.\ \cite{Kim:08} was used to represent 
   the protein-protein interaction potential in Eq.\ (\ref{eq:18}): 
   $\sigma_0=31.8$ \AA, 
   $\epsilon/k_{\text{B}}= 419$ K, $\delta = 4.14$ \AA. }
 \footnotetext[2]{Calculated from equilibrated protein geometry 
   and atomic partial charges \cite{LeitnerComm}.}
 \footnotetext[3]{According to MD simulation data from Ref.\
   \cite{DMpre2:08}. }
 \footnotetext[4]{From the vdW volume of the protein using the Amber
   FF03 force field. }
 \footnotetext[5]{$\langle m_0\rangle$ calculated from the MD trajectory 
   relative the center of mass, total charge
   of the Ox state of the protein is $-8$. Fluctuations of the protein
   dipole are caused by protein's vibrations. }
 \footnotetext[6]{Calculated from the exponential fit of the time
   self-correlation function of the protein
   dipole.  } 
\end{table}

The calculation of the solute dipole component of the dielectric
response simplifies in the continuum limit when the integral in Eq.\
(\ref{eq:20}) loses the dependence on frequency and reduces to Eq.\
(\ref{eq:27}). This integral depends on two parameters, the volume
fraction $\eta_0$ and the reduced geometry parameter $r=1/2+\sigma/ (2\sigma_0)$,
when the hard-spheres approximation is used for the density structure
factor $S_0(k,\eta_0)$. The range $0.5 \leq r \leq1 $ covers most problems of
interest. Numerical integration of Eq.\ (\ref{eq:27}) with the PY
density structure factor \cite{Hansen:03} was done in this range of
$r$-values and volume fractions in the range $0\leq \eta_0\leq 0.3$. The
numerical results were interpolated with the polynomial function
\begin{equation}
  \label{eqA:3}
  I(\eta_0,r) = a(\eta_0) + b(\eta_0) r^2 + c(\eta_0) r^4 + d(\eta_0) r^6,
\end{equation}
where
\begin{equation}
  \label{eqA:4}
  \begin{split}
    a(\eta_0) & = 1 + 0.225 \eta_0 + 7.726 \eta_0^2 - 13.805 \eta_0^3\\
    b(\eta_0) & = -9.694 \eta_0  - 18.572 \eta_0^2 + 16.642 \eta_0^3 \\
    c(\eta_0) & = 6.987 \eta_0 + 38.913 \eta_0^2 - 5.940 \eta_0^3 \\
    d(\eta_0) & = 2.108 \eta_0 - 16.570 \eta_0^2 - 10.007 \eta_0^3\\
  \end{split}
\end{equation}
The expansion in even powers in $r$ in Eq.\ (\ref{eqA:3}) is dictated
by the symmetry of the density structure factor \cite{Hansen:03}, and
the density expansion of the polynomial coefficients has been chosen
to justify the ideal-solution limit $I(0,r)=1$.

\bibliographystyle{apsrev}
\bibliography{chem_abbr,dielectric,dm,statmech,protein,liquids,solvation,dynamics}

\begin{thebibliography}{53}
\expandafter\ifx\csname natexlab\endcsname\relax\def\natexlab#1{#1}\fi
\expandafter\ifx\csname bibnamefont\endcsname\relax
  \def\bibnamefont#1{#1}\fi
\expandafter\ifx\csname bibfnamefont\endcsname\relax
  \def\bibfnamefont#1{#1}\fi
\expandafter\ifx\csname citenamefont\endcsname\relax
  \def\citenamefont#1{#1}\fi
\expandafter\ifx\csname url\endcsname\relax
  \def\url#1{\texttt{#1}}\fi
\expandafter\ifx\csname urlprefix\endcsname\relax\def\urlprefix{URL }\fi
\providecommand{\bibinfo}[2]{#2}
\providecommand{\eprint}[2][]{\url{#2}}

\bibitem[{\citenamefont{Scaife}(1998)}]{Scaife:98}
\bibinfo{author}{\bibfnamefont{B.~K.~P.} \bibnamefont{Scaife}},
  \emph{\bibinfo{title}{Principles of dilectrics}}
  (\bibinfo{publisher}{Clarendon Press}, \bibinfo{address}{Oxford},
  \bibinfo{year}{1998}).

\bibitem[{\citenamefont{Choi}(1999)}]{Choi:99}
\bibinfo{author}{\bibfnamefont{T.~C.} \bibnamefont{Choi}},
  \emph{\bibinfo{title}{Effective Medium Theory}}
  (\bibinfo{publisher}{Clarendon Press}, \bibinfo{address}{Oxford},
  \bibinfo{year}{1999}).

\bibitem[{\citenamefont{Takashima}(1989)}]{Takashima:89}
\bibinfo{author}{\bibfnamefont{S.}~\bibnamefont{Takashima}},
  \emph{\bibinfo{title}{Electrical properties of biopolymers and membranes}}
  (\bibinfo{publisher}{Adam Hilger}, \bibinfo{address}{Bristol},
  \bibinfo{year}{1989}).

\bibitem[{\citenamefont{Beard et~al.}(2002)\citenamefont{Beard, Turner, and
  Schmuttenmaer}}]{Beard:02}
\bibinfo{author}{\bibfnamefont{M.~C.} \bibnamefont{Beard}},
  \bibinfo{author}{\bibfnamefont{G.~M.} \bibnamefont{Turner}},
  \bibnamefont{and} \bibinfo{author}{\bibfnamefont{C.~A.}
  \bibnamefont{Schmuttenmaer}}, \bibinfo{journal}{J. Phys. Chem. B}
  \textbf{\bibinfo{volume}{106}}, \bibinfo{pages}{7146} (\bibinfo{year}{2002}).

\bibitem[{\citenamefont{Yokoyama et~al.}(2001)\citenamefont{Yokoyama, Kamei,
  Minami, and Suzuki}}]{Yokoyama:01}
\bibinfo{author}{\bibfnamefont{K.}~\bibnamefont{Yokoyama}},
  \bibinfo{author}{\bibfnamefont{T.}~\bibnamefont{Kamei}},
  \bibinfo{author}{\bibfnamefont{H.}~\bibnamefont{Minami}}, \bibnamefont{and}
  \bibinfo{author}{\bibfnamefont{M.}~\bibnamefont{Suzuki}},
  \bibinfo{journal}{J. Phys. Chem. B} \textbf{\bibinfo{volume}{105}},
  \bibinfo{pages}{12622} (\bibinfo{year}{2001}).

\bibitem[{\citenamefont{Bergner et~al.}(2005)\citenamefont{Bergner, Heugen,
  Br\"{u}ndermann, Schwaab, Havenith, Chamberlin, and Haller}}]{Bergner:05}
\bibinfo{author}{\bibfnamefont{A.}~\bibnamefont{Bergner}},
  \bibinfo{author}{\bibfnamefont{U.}~\bibnamefont{Heugen}},
  \bibinfo{author}{\bibfnamefont{E.}~\bibnamefont{Br\"{u}ndermann}},
  \bibinfo{author}{\bibfnamefont{G.}~\bibnamefont{Schwaab}},
  \bibinfo{author}{\bibfnamefont{M.}~\bibnamefont{Havenith}},
  \bibinfo{author}{\bibfnamefont{D.~R.} \bibnamefont{Chamberlin}},
  \bibnamefont{and} \bibinfo{author}{\bibfnamefont{E.~E.}
  \bibnamefont{Haller}}, \bibinfo{journal}{Rev.\ Sci.\ Instrum.}
  \textbf{\bibinfo{volume}{76}}, \bibinfo{eid}{063110} (\bibinfo{year}{2005}).

\bibitem[{\citenamefont{Zhang and Durbin}(2006)}]{Zhang:06}
\bibinfo{author}{\bibfnamefont{C.}~\bibnamefont{Zhang}} \bibnamefont{and}
  \bibinfo{author}{\bibfnamefont{S.~M.} \bibnamefont{Durbin}},
  \bibinfo{journal}{J. Phys. Chem. B} \textbf{\bibinfo{volume}{110}},
  \bibinfo{pages}{23607} (\bibinfo{year}{2006}).

\bibitem[{\citenamefont{Xu et~al.}(2006)\citenamefont{Xu, Plaxco, and
  Allen}}]{Xu:06}
\bibinfo{author}{\bibfnamefont{J.}~\bibnamefont{Xu}},
  \bibinfo{author}{\bibfnamefont{K.~W.} \bibnamefont{Plaxco}},
  \bibnamefont{and} \bibinfo{author}{\bibfnamefont{S.~J.} \bibnamefont{Allen}},
  \bibinfo{journal}{Prot. Science} \textbf{\bibinfo{volume}{15}},
  \bibinfo{pages}{1175} (\bibinfo{year}{2006}).

\bibitem[{\citenamefont{Knab et~al.}(2007)\citenamefont{Knab, Chen, He, and
  Markelz}}]{Knab:07}
\bibinfo{author}{\bibfnamefont{J.~R.} \bibnamefont{Knab}},
  \bibinfo{author}{\bibfnamefont{J.-Y.} \bibnamefont{Chen}},
  \bibinfo{author}{\bibfnamefont{Y.}~\bibnamefont{He}}, \bibnamefont{and}
  \bibinfo{author}{\bibfnamefont{A.~G.} \bibnamefont{Markelz}},
  \bibinfo{journal}{Proc. IEEE} \textbf{\bibinfo{volume}{95}},
  \bibinfo{pages}{1605} (\bibinfo{year}{2007}).

\bibitem[{\citenamefont{Ebbinghaus et~al.}(2008)\citenamefont{Ebbinghaus, Kim,
  Heyden, Yu, Gruebele, Leitner, and Havenith}}]{Ebbinghaus:08}
\bibinfo{author}{\bibfnamefont{S.}~\bibnamefont{Ebbinghaus}},
  \bibinfo{author}{\bibfnamefont{S.~J.} \bibnamefont{Kim}},
  \bibinfo{author}{\bibfnamefont{M.}~\bibnamefont{Heyden}},
  \bibinfo{author}{\bibfnamefont{X.}~\bibnamefont{Yu}},
  \bibinfo{author}{\bibfnamefont{M.}~\bibnamefont{Gruebele}},
  \bibinfo{author}{\bibfnamefont{D.~M.} \bibnamefont{Leitner}},
  \bibnamefont{and} \bibinfo{author}{\bibfnamefont{M.}~\bibnamefont{Havenith}},
  \bibinfo{journal}{J. Am. Chem. Soc.} \textbf{\bibinfo{volume}{130}},
  \bibinfo{pages}{2374} (\bibinfo{year}{2008}).

\bibitem[{\citenamefont{Frauenfelder et~al.}(2009)\citenamefont{Frauenfelder,
  Chen, Berendzen, Fenimore, Jansson, McMahon, Stroe, Swenson, and
  Young}}]{Frauenfelder:09}
\bibinfo{author}{\bibfnamefont{H.}~\bibnamefont{Frauenfelder}},
  \bibinfo{author}{\bibfnamefont{G.}~\bibnamefont{Chen}},
  \bibinfo{author}{\bibfnamefont{J.}~\bibnamefont{Berendzen}},
  \bibinfo{author}{\bibfnamefont{P.~W.} \bibnamefont{Fenimore}},
  \bibinfo{author}{\bibfnamefont{H.}~\bibnamefont{Jansson}},
  \bibinfo{author}{\bibfnamefont{B.~H.} \bibnamefont{McMahon}},
  \bibinfo{author}{\bibfnamefont{I.~R.} \bibnamefont{Stroe}},
  \bibinfo{author}{\bibfnamefont{J.}~\bibnamefont{Swenson}}, \bibnamefont{and}
  \bibinfo{author}{\bibfnamefont{R.~D.} \bibnamefont{Young}},
  \bibinfo{journal}{Proc. Nat. Acad. Sci. USA} \textbf{\bibinfo{volume}{106}},
  \bibinfo{pages}{5129} (\bibinfo{year}{2009}).

\bibitem[{\citenamefont{Baxter and Schmuttenmaer}(2006)}]{Baxter:06}
\bibinfo{author}{\bibfnamefont{J.~B.} \bibnamefont{Baxter}} \bibnamefont{and}
  \bibinfo{author}{\bibfnamefont{C.~A.} \bibnamefont{Schmuttenmaer}},
  \bibinfo{journal}{J. Phys. Chem. B} \textbf{\bibinfo{volume}{110}},
  \bibinfo{pages}{25229} (\bibinfo{year}{2006}).

\bibitem[{\citenamefont{Alcoutlabi and McKenna}(2005)}]{Alcoutlabi:05}
\bibinfo{author}{\bibfnamefont{M.}~\bibnamefont{Alcoutlabi}} \bibnamefont{and}
  \bibinfo{author}{\bibfnamefont{G.~B.} \bibnamefont{McKenna}},
  \bibinfo{journal}{J. Phys.: Condens. Matter} \textbf{\bibinfo{volume}{17}},
  \bibinfo{pages}{R461} (\bibinfo{year}{2005}).

\bibitem[{\citenamefont{Matyushov}(2004{\natexlab{a}})}]{DMjcp1:04}
\bibinfo{author}{\bibfnamefont{D.~V.} \bibnamefont{Matyushov}},
  \bibinfo{journal}{J. Chem. Phys.} \textbf{\bibinfo{volume}{120}},
  \bibinfo{pages}{1375} (\bibinfo{year}{2004}{\natexlab{a}}).

\bibitem[{\citenamefont{LeBard and Matyushov}(2008{\natexlab{a}})}]{DMjcp2:08}
\bibinfo{author}{\bibfnamefont{D.~N.} \bibnamefont{LeBard}} \bibnamefont{and}
  \bibinfo{author}{\bibfnamefont{D.~V.} \bibnamefont{Matyushov}},
  \bibinfo{journal}{J. Chem. Phys.} \textbf{\bibinfo{volume}{128}},
  \bibinfo{pages}{155106} (\bibinfo{year}{2008}{\natexlab{a}}).

\bibitem[{\citenamefont{Heyden et~al.}(2008)\citenamefont{Heyden,
  Br{\"u}ndermann, Heugen, Leitner, and Havenith}}]{Heyden:08}
\bibinfo{author}{\bibfnamefont{M.}~\bibnamefont{Heyden}},
  \bibinfo{author}{\bibfnamefont{E.}~\bibnamefont{Br{\"u}ndermann}},
  \bibinfo{author}{\bibfnamefont{U.}~\bibnamefont{Heugen}},
  \bibinfo{author}{\bibfnamefont{D.~M.} \bibnamefont{Leitner}},
  \bibnamefont{and} \bibinfo{author}{\bibfnamefont{M.}~\bibnamefont{Havenith}},
  \bibinfo{journal}{J. Am. Chem. Soc.} \textbf{\bibinfo{volume}{130}},
  \bibinfo{pages}{5773} (\bibinfo{year}{2008}).

\bibitem[{\citenamefont{Ebbinghaus et~al.}(2007)\citenamefont{Ebbinghaus, Kim,
  Heyden, Yu, Heugen, Gruebele, Leitner, and Havenith}}]{Ebbinghaus:07}
\bibinfo{author}{\bibfnamefont{S.}~\bibnamefont{Ebbinghaus}},
  \bibinfo{author}{\bibfnamefont{S.~J.} \bibnamefont{Kim}},
  \bibinfo{author}{\bibfnamefont{M.}~\bibnamefont{Heyden}},
  \bibinfo{author}{\bibfnamefont{X.}~\bibnamefont{Yu}},
  \bibinfo{author}{\bibfnamefont{U.}~\bibnamefont{Heugen}},
  \bibinfo{author}{\bibfnamefont{M.}~\bibnamefont{Gruebele}},
  \bibinfo{author}{\bibfnamefont{D.~M.} \bibnamefont{Leitner}},
  \bibnamefont{and} \bibinfo{author}{\bibfnamefont{M.}~\bibnamefont{Havenith}},
  \bibinfo{journal}{Proc. Nat. Acad. Sci. USA} \textbf{\bibinfo{volume}{104}},
  \bibinfo{pages}{20749} (\bibinfo{year}{2007}).

\bibitem[{\citenamefont{LeBard and Matyushov}(2008{\natexlab{b}})}]{DMpre2:08}
\bibinfo{author}{\bibfnamefont{D.~N.} \bibnamefont{LeBard}} \bibnamefont{and}
  \bibinfo{author}{\bibfnamefont{D.~V.} \bibnamefont{Matyushov}},
  \bibinfo{journal}{Phys. Rev. E} \textbf{\bibinfo{volume}{78}},
  \bibinfo{pages}{061901} (\bibinfo{year}{2008}{\natexlab{b}}).

\bibitem[{\citenamefont{LeBard and Matyushov}(2009)}]{DMjpcb2:09}
\bibinfo{author}{\bibfnamefont{D.~N.} \bibnamefont{LeBard}} \bibnamefont{and}
  \bibinfo{author}{\bibfnamefont{D.~V.} \bibnamefont{Matyushov}},
  \bibinfo{journal}{J. Phys. Chem. B} p. \bibinfo{pages}{to be submitted}
  (\bibinfo{year}{2009}).

\bibitem[{\citenamefont{Landau and Lifshitz}(1984)}]{Landau8}
\bibinfo{author}{\bibfnamefont{L.~D.} \bibnamefont{Landau}} \bibnamefont{and}
  \bibinfo{author}{\bibfnamefont{E.~M.} \bibnamefont{Lifshitz}},
  \emph{\bibinfo{title}{Electrodynamics of continuous media}}
  (\bibinfo{publisher}{Pergamon}, \bibinfo{address}{Oxford},
  \bibinfo{year}{1984}).

\bibitem[{\citenamefont{B{{\"o}}ttcher}(1973)}]{Boettcher:73}
\bibinfo{author}{\bibfnamefont{C.~J.~F.} \bibnamefont{B{{\"o}}ttcher}},
  \emph{\bibinfo{title}{Theory of Electric Polarization}},
  vol.~\bibinfo{volume}{1} (\bibinfo{publisher}{Elsevier},
  \bibinfo{address}{Amsterdam}, \bibinfo{year}{1973}).

\bibitem[{\citenamefont{Madden and Kivelson}(1984)}]{Madden:84}
\bibinfo{author}{\bibfnamefont{P.}~\bibnamefont{Madden}} \bibnamefont{and}
  \bibinfo{author}{\bibfnamefont{D.}~\bibnamefont{Kivelson}},
  \bibinfo{journal}{Adv. Chem. Phys.} \textbf{\bibinfo{volume}{56}},
  \bibinfo{pages}{467} (\bibinfo{year}{1984}).

\bibitem[{\citenamefont{Neumann}(1986)}]{Neumann:86}
\bibinfo{author}{\bibfnamefont{M.}~\bibnamefont{Neumann}},
  \bibinfo{journal}{Mol. Phys.} \textbf{\bibinfo{volume}{57}},
  \bibinfo{pages}{97} (\bibinfo{year}{1986}).

\bibitem[{\citenamefont{Chandler}(1993)}]{Chandler:93}
\bibinfo{author}{\bibfnamefont{D.}~\bibnamefont{Chandler}},
  \bibinfo{journal}{Phys. Rev. E} \textbf{\bibinfo{volume}{48}},
  \bibinfo{pages}{2898} (\bibinfo{year}{1993}).

\bibitem[{\citenamefont{Hansen and McDonald}(2003)}]{Hansen:03}
\bibinfo{author}{\bibfnamefont{J.~P.} \bibnamefont{Hansen}} \bibnamefont{and}
  \bibinfo{author}{\bibfnamefont{I.~R.} \bibnamefont{McDonald}},
  \emph{\bibinfo{title}{Theory of Simple Liquids}}
  (\bibinfo{publisher}{Academic Press}, \bibinfo{address}{Amsterdam},
  \bibinfo{year}{2003}).

\bibitem[{\citenamefont{Wertheim}(1971)}]{Wertheim:71}
\bibinfo{author}{\bibfnamefont{M.~S.} \bibnamefont{Wertheim}},
  \bibinfo{journal}{J. Chem. Phys.} \textbf{\bibinfo{volume}{55}},
  \bibinfo{pages}{4291} (\bibinfo{year}{1971}).

\bibitem[{\citenamefont{Matyushov}(2005)}]{DMjcp1:05}
\bibinfo{author}{\bibfnamefont{D.~V.} \bibnamefont{Matyushov}},
  \bibinfo{journal}{J. Chem. Phys.} \textbf{\bibinfo{volume}{122}},
  \bibinfo{pages}{044502} (\bibinfo{year}{2005}).

\bibitem[{\citenamefont{Stell et~al.}(1981)\citenamefont{Stell, Patey, and
  H{{\o}}ye}}]{SPH:81}
\bibinfo{author}{\bibfnamefont{G.}~\bibnamefont{Stell}},
  \bibinfo{author}{\bibfnamefont{G.~N.} \bibnamefont{Patey}}, \bibnamefont{and}
  \bibinfo{author}{\bibfnamefont{J.~S.} \bibnamefont{H{{\o}}ye}},
  \bibinfo{journal}{Adv. Chem. Phys.} \textbf{\bibinfo{volume}{48}},
  \bibinfo{pages}{183} (\bibinfo{year}{1981}).

\bibitem[{\citenamefont{Onsager}(1936)}]{Onsager:36}
\bibinfo{author}{\bibfnamefont{L.}~\bibnamefont{Onsager}}, \bibinfo{journal}{J.
  Am. Chem. Soc.} \textbf{\bibinfo{volume}{58}}, \bibinfo{pages}{1486}
  (\bibinfo{year}{1936}).

\bibitem[{\citenamefont{Martin and Matyushov}(2008)}]{DMepl:08}
\bibinfo{author}{\bibfnamefont{D.~R.} \bibnamefont{Martin}} \bibnamefont{and}
  \bibinfo{author}{\bibfnamefont{D.~V.} \bibnamefont{Matyushov}},
  \bibinfo{journal}{Europhys.\ Lett.} \textbf{\bibinfo{volume}{82}},
  \bibinfo{pages}{16003} (\bibinfo{year}{2008}).

\bibitem[{\citenamefont{Kirkwood and Buff}(1951)}]{Kirkwood:51}
\bibinfo{author}{\bibfnamefont{J.~G.} \bibnamefont{Kirkwood}} \bibnamefont{and}
  \bibinfo{author}{\bibfnamefont{F.~B.} \bibnamefont{Buff}},
  \bibinfo{journal}{J. Chem. Phys.} \textbf{\bibinfo{volume}{19}},
  \bibinfo{pages}{774} (\bibinfo{year}{1951}).

\bibitem[{\citenamefont{Belloni}(2000)}]{Belloni:00}
\bibinfo{author}{\bibfnamefont{L.}~\bibnamefont{Belloni}}, \bibinfo{journal}{J.
  Phys.: Condens. Matter} \textbf{\bibinfo{volume}{12}}, \bibinfo{pages}{R549}
  (\bibinfo{year}{2000}).

\bibitem[{\citenamefont{Matyushov}(2004{\natexlab{b}})}]{DMjcp2:04}
\bibinfo{author}{\bibfnamefont{D.~V.} \bibnamefont{Matyushov}},
  \bibinfo{journal}{J. Chem. Phys.} \textbf{\bibinfo{volume}{120}},
  \bibinfo{pages}{7532} (\bibinfo{year}{2004}{\natexlab{b}}).

\bibitem[{\citenamefont{Bagchi and Chandra}(1991)}]{Bagchi:91}
\bibinfo{author}{\bibfnamefont{B.}~\bibnamefont{Bagchi}} \bibnamefont{and}
  \bibinfo{author}{\bibfnamefont{A.}~\bibnamefont{Chandra}},
  \bibinfo{journal}{Adv. Chem. Phys.} \textbf{\bibinfo{volume}{80}},
  \bibinfo{pages}{1} (\bibinfo{year}{1991}).

\bibitem[{\citenamefont{Kihara and Miyoshi}(1975)}]{Kihara:75}
\bibinfo{author}{\bibfnamefont{T.}~\bibnamefont{Kihara}} \bibnamefont{and}
  \bibinfo{author}{\bibfnamefont{K.}~\bibnamefont{Miyoshi}},
  \bibinfo{journal}{J. Stat. Phys.} \textbf{\bibinfo{volume}{13}},
  \bibinfo{pages}{337} (\bibinfo{year}{1975}).

\bibitem[{\citenamefont{Barrio and Solana}(1999)}]{Barrio:99}
\bibinfo{author}{\bibfnamefont{C.}~\bibnamefont{Barrio}} \bibnamefont{and}
  \bibinfo{author}{\bibfnamefont{J.~R.} \bibnamefont{Solana}},
  \bibinfo{journal}{Mol. Phys.} \textbf{\bibinfo{volume}{97}},
  \bibinfo{pages}{797} (\bibinfo{year}{1999}).

\bibitem[{\citenamefont{Zhang et~al.}(2007)\citenamefont{Zhang, Wang, Kao, Qiu,
  Yang, Okobiah, and Zhong}}]{Zhang:07}
\bibinfo{author}{\bibfnamefont{L.}~\bibnamefont{Zhang}},
  \bibinfo{author}{\bibfnamefont{L.}~\bibnamefont{Wang}},
  \bibinfo{author}{\bibfnamefont{Y.-T.} \bibnamefont{Kao}},
  \bibinfo{author}{\bibfnamefont{W.}~\bibnamefont{Qiu}},
  \bibinfo{author}{\bibfnamefont{Y.}~\bibnamefont{Yang}},
  \bibinfo{author}{\bibfnamefont{O.}~\bibnamefont{Okobiah}}, \bibnamefont{and}
  \bibinfo{author}{\bibfnamefont{D.}~\bibnamefont{Zhong}},
  \bibinfo{journal}{Proc. Nat. Acad. Sci. USA} \textbf{\bibinfo{volume}{104}},
  \bibinfo{pages}{18461} (\bibinfo{year}{2007}).

\bibitem[{\citenamefont{Shukla et~al.}(2008)\citenamefont{Shukla, Mylonas,
  Cola, Finet, Timmins, Narayanan, and Svergun}}]{Shukla:08}
\bibinfo{author}{\bibfnamefont{A.}~\bibnamefont{Shukla}},
  \bibinfo{author}{\bibfnamefont{E.}~\bibnamefont{Mylonas}},
  \bibinfo{author}{\bibfnamefont{E.~D.} \bibnamefont{Cola}},
  \bibinfo{author}{\bibfnamefont{S.}~\bibnamefont{Finet}},
  \bibinfo{author}{\bibfnamefont{P.}~\bibnamefont{Timmins}},
  \bibinfo{author}{\bibfnamefont{T.}~\bibnamefont{Narayanan}},
  \bibnamefont{and} \bibinfo{author}{\bibfnamefont{D.~I.}
  \bibnamefont{Svergun}}, \bibinfo{journal}{Proc. Nat. Acad. Sci. USA}
  \textbf{\bibinfo{volume}{105}}, \bibinfo{pages}{5075} (\bibinfo{year}{2008}).

\bibitem[{\citenamefont{Liu et~al.}(2005)\citenamefont{Liu, Chen, and
  Chen}}]{Liu:05}
\bibinfo{author}{\bibfnamefont{Y.}~\bibnamefont{Liu}},
  \bibinfo{author}{\bibfnamefont{W.-R.} \bibnamefont{Chen}}, \bibnamefont{and}
  \bibinfo{author}{\bibfnamefont{S.-H.} \bibnamefont{Chen}},
  \bibinfo{journal}{J. Chem. Phys.} \textbf{\bibinfo{volume}{122}},
  \bibinfo{pages}{044507} (\bibinfo{year}{2005}).

\bibitem[{\citenamefont{Croxton}(1975)}]{Croxton:75}
\bibinfo{author}{\bibfnamefont{C.~A.} \bibnamefont{Croxton}},
  \emph{\bibinfo{title}{Introduction to liquid state physics}}
  (\bibinfo{publisher}{Wiley}, \bibinfo{address}{New York},
  \bibinfo{year}{1975}).

\bibitem[{\citenamefont{Rudas et~al.}(2006)\citenamefont{Rudas, Schr\"{o}der,
  Boresch, and Steinhauser}}]{Rudas2:06}
\bibinfo{author}{\bibfnamefont{T.}~\bibnamefont{Rudas}},
  \bibinfo{author}{\bibfnamefont{C.}~\bibnamefont{Schr\"{o}der}},
  \bibinfo{author}{\bibfnamefont{S.}~\bibnamefont{Boresch}}, \bibnamefont{and}
  \bibinfo{author}{\bibfnamefont{O.}~\bibnamefont{Steinhauser}},
  \bibinfo{journal}{J. Chem. Phys.} \textbf{\bibinfo{volume}{124}},
  \bibinfo{pages}{234908} (\bibinfo{year}{2006}).

\bibitem[{\citenamefont{Wang}(1985)}]{WangCom:85}
\bibinfo{author}{\bibfnamefont{C.~H.} \bibnamefont{Wang}},
  \emph{\bibinfo{title}{Spectroscopy of Condensed Media. Dynamics of Molecular
  Interactions}} (\bibinfo{publisher}{Acedemic Press},
  \bibinfo{address}{Orlando}, \bibinfo{year}{1985}), \bibinfo{note}{there is a
  typo in Eq.\ (1.175) in the book, the dielectric constant $\epsilon'(\omega)$
  should be under square root, as in Eq.\ (22).}

\bibitem[{\citenamefont{McQuarrie}(2000)}]{McQuarrie:00}
\bibinfo{author}{\bibfnamefont{D.~A.} \bibnamefont{McQuarrie}},
  \emph{\bibinfo{title}{Statistical Mechanics}} (\bibinfo{publisher}{University
  Science Books}, \bibinfo{address}{Sausalito, CA}, \bibinfo{year}{2000}).

\bibitem[{\citenamefont{Yada et~al.}(2009)\citenamefont{Yada, Nagai, and
  Tanaka}}]{Yada:09}
\bibinfo{author}{\bibfnamefont{H.}~\bibnamefont{Yada}},
  \bibinfo{author}{\bibfnamefont{M.}~\bibnamefont{Nagai}}, \bibnamefont{and}
  \bibinfo{author}{\bibfnamefont{K.}~\bibnamefont{Tanaka}},
  \bibinfo{journal}{Chem. Phys. Lett.} \textbf{\bibinfo{volume}{473}},
  \bibinfo{pages}{279} (\bibinfo{year}{2009}).

\bibitem[{\citenamefont{Gharsallaoui et~al.}(2008)\citenamefont{Gharsallaoui,
  Roge, Genotelle, and Mathlouthi}}]{Gharsallaoui:08}
\bibinfo{author}{\bibfnamefont{A.}~\bibnamefont{Gharsallaoui}},
  \bibinfo{author}{\bibfnamefont{B.}~\bibnamefont{Roge}},
  \bibinfo{author}{\bibfnamefont{J.}~\bibnamefont{Genotelle}},
  \bibnamefont{and}
  \bibinfo{author}{\bibfnamefont{M.}~\bibnamefont{Mathlouthi}},
  \bibinfo{journal}{Food Chem.} \textbf{\bibinfo{volume}{106}},
  \bibinfo{pages}{1443} (\bibinfo{year}{2008}).

\bibitem[{\citenamefont{Yand and Gruebele}(2003)}]{YangNature:03}
\bibinfo{author}{\bibfnamefont{W.~Y.} \bibnamefont{Yand}} \bibnamefont{and}
  \bibinfo{author}{\bibfnamefont{M.}~\bibnamefont{Gruebele}},
  \bibinfo{journal}{Nature} \textbf{\bibinfo{volume}{423}},
  \bibinfo{pages}{193} (\bibinfo{year}{2003}).

\bibitem[{\citenamefont{Zhang et~al.}(2008)\citenamefont{Zhang, Skoda, Jacobs,
  Martin, Martin, and Schreiber}}]{Zhang:08}
\bibinfo{author}{\bibfnamefont{F.}~\bibnamefont{Zhang}},
  \bibinfo{author}{\bibfnamefont{M.~W.~A.} \bibnamefont{Skoda}},
  \bibinfo{author}{\bibfnamefont{R.~M.~J.} \bibnamefont{Jacobs}},
  \bibinfo{author}{\bibfnamefont{R.~A.} \bibnamefont{Martin}},
  \bibinfo{author}{\bibfnamefont{C.~M.} \bibnamefont{Martin}},
  \bibnamefont{and}
  \bibinfo{author}{\bibfnamefont{F.}~\bibnamefont{Schreiber}},
  \bibinfo{journal}{J. Phys. Chem. B} \textbf{\bibinfo{volume}{111}},
  \bibinfo{pages}{251} (\bibinfo{year}{2008}).

\bibitem[{\citenamefont{Kim et~al.}(2008)\citenamefont{Kim, Dumont, and
  Gruebele}}]{Kim:08}
\bibinfo{author}{\bibfnamefont{S.~J.} \bibnamefont{Kim}},
  \bibinfo{author}{\bibfnamefont{C.}~\bibnamefont{Dumont}}, \bibnamefont{and}
  \bibinfo{author}{\bibfnamefont{M.}~\bibnamefont{Gruebele}},
  \bibinfo{journal}{Biophys. J.} \textbf{\bibinfo{volume}{94}},
  \bibinfo{pages}{4924} (\bibinfo{year}{2008}).

\bibitem[{\citenamefont{Javid et~al.}(2007)\citenamefont{Javid, Voggt, Krywka,
  Tolan, and Winter}}]{Javid:07}
\bibinfo{author}{\bibfnamefont{N.}~\bibnamefont{Javid}},
  \bibinfo{author}{\bibfnamefont{K.}~\bibnamefont{Voggt}},
  \bibinfo{author}{\bibfnamefont{C.}~\bibnamefont{Krywka}},
  \bibinfo{author}{\bibfnamefont{M.}~\bibnamefont{Tolan}}, \bibnamefont{and}
  \bibinfo{author}{\bibfnamefont{R.}~\bibnamefont{Winter}},
  \bibinfo{journal}{ChemPhysChem} \textbf{\bibinfo{volume}{8}},
  \bibinfo{pages}{679} (\bibinfo{year}{2007}).

\bibitem[{\citenamefont{Weing\"{a}rtner
  et~al.}(2001)\citenamefont{Weing\"{a}rtner, Knocks, Boresch, H\"{o}chtl, and
  Steinhauser}}]{Weingartner:01}
\bibinfo{author}{\bibfnamefont{H.}~\bibnamefont{Weing\"{a}rtner}},
  \bibinfo{author}{\bibfnamefont{A.}~\bibnamefont{Knocks}},
  \bibinfo{author}{\bibfnamefont{S.}~\bibnamefont{Boresch}},
  \bibinfo{author}{\bibfnamefont{P.}~\bibnamefont{H\"{o}chtl}},
  \bibnamefont{and}
  \bibinfo{author}{\bibfnamefont{O.}~\bibnamefont{Steinhauser}},
  \bibinfo{journal}{J.\ Chem.\ Phys.} \textbf{\bibinfo{volume}{115}},
  \bibinfo{pages}{1463} (\bibinfo{year}{2001}).

\bibitem[{\citenamefont{Lee et~al.}(2005)\citenamefont{Lee, Debenedetti, and
  Errington}}]{Lee:05}
\bibinfo{author}{\bibfnamefont{S.~L.} \bibnamefont{Lee}},
  \bibinfo{author}{\bibfnamefont{P.~G.} \bibnamefont{Debenedetti}},
  \bibnamefont{and} \bibinfo{author}{\bibfnamefont{J.~R.}
  \bibnamefont{Errington}}, \bibinfo{journal}{J. Chem. Phys.}
  \textbf{\bibinfo{volume}{122}}, \bibinfo{eid}{204511} (\bibinfo{year}{2005}).

\bibitem[{\citenamefont{Paolantoni et~al.}(2009)\citenamefont{Paolantoni,
  Comez, Gallina, Sassi, Scarponi, Fioretto, and Morresi}}]{Paolantoni:09}
\bibinfo{author}{\bibfnamefont{M.}~\bibnamefont{Paolantoni}},
  \bibinfo{author}{\bibfnamefont{L.}~\bibnamefont{Comez}},
  \bibinfo{author}{\bibfnamefont{M.~E.} \bibnamefont{Gallina}},
  \bibinfo{author}{\bibfnamefont{P.}~\bibnamefont{Sassi}},
  \bibinfo{author}{\bibfnamefont{F.}~\bibnamefont{Scarponi}},
  \bibinfo{author}{\bibfnamefont{D.}~\bibnamefont{Fioretto}}, \bibnamefont{and}
  \bibinfo{author}{\bibfnamefont{A.}~\bibnamefont{Morresi}},
  \bibinfo{journal}{J. Phys. Chem. B} \textbf{\bibinfo{volume}{113}},
  \bibinfo{pages}{7874} (\bibinfo{year}{2009}).

\bibitem[{\citenamefont{Leitner}()}]{LeitnerComm}
\bibinfo{author}{\bibfnamefont{D.~M.} \bibnamefont{Leitner}},
  \bibinfo{note}{private communication}.

\end{thebibliography}

\end{document}